\documentstyle[aps,epsfig,floats,prl]{revtex} 
\begin{document}
\begin{twocolumn}
\draft
\wideabs{
\title{Two-Phase Regime in the Magnetic Field-Temperature Phase Diagram of a
Type-II Superconductor}
\author{L. L. A. Adams, Klaus Halterman, Oriol T. Valls, and A. M. Goldman}
\address{School of Physics and Astronomy, University of Minnesota, 
116 Church St. SE,%
\\
Minneapolis,
Minnesota 55455}
\date{%TCIMACRO{\TeXButton{TeX field}{\today} }
%BeginExpansion
\today%
%EndExpansion
}
\maketitle

\begin{abstract}
The magnetic field and temperature dependencies of the magnetic moments of
superconducting crystals of ${\rm V_{3}Si}$ have been studied. In a constant
magnetic field and at temperatures somewhat below the superconducting
transition temperature, the moments are hysteretic in temperature. However
the magnetic moment-magnetic field isotherms are reversible and exhibit
features that formally resemble the pressure-volume isotherms of the
liquid-gas transition. This suggests the existence of a first-order phase
transition, a two-phase regime, and a critical point in the superconducting
phase diagram. The entropy change, determined from the data using the
Clausius-Clapeyron equation, is consistent with estimates based on the
difference in the vortex densities of the two phases.
\end{abstract}

}
The study of magnetic field-temperature phase diagrams of type-II
superconductors has revealed a rich range of behaviors, with the various
transitions between vortex configurations, i.e., from lattice to liquid,
glass to liquid, etc., resembling those found in ordinary matter. As a
consequence, the study of thermodynamic properties of ``vortex matter'' has
emerged as an important field of research\cite{Crabtree}. The richness of
the physics became apparent only after high-quality single-crystal samples
of high temperature superconductors became available \cite{Larkin}.
Conventional superconductors that were subsequently studied, were also found
to exhibit related phenomena, examples of which are phase transitions \cite
{Ling} associated with the peak effect \cite{Pippard} in Nb, and coexistent
vortex phases in the peak effect regime of NbSe$_{2}$\cite{Marchevsky}. The
work on Nb combined magnetic measurements with small angle neutron
scattering studies, and that on NbSe$_{2}$ involved a direct map of the
local ac induction using a variant of scanning Hall-probe microscopy.
Domains of differing pinning strength remained distinct during the
transformation from stronger pinning and higher critical current at high
temperature to weaker pinning and lower critical current at low temperature.
Such behavior is a hallmark of a first order transition. In this Letter we
report measurements of the equilibrium magnetic moments of V$_{3}$Si
crystals as a function of magnetic field and temperature. We provide an
interpretation of the results using an analogy with the phase diagram of the
liquid-gas system that further supports the existence of coexistent phases
in addition to the existence of a critical point. Our approach, being based
on a thermodynamic analysis of magnetization data alone, does not reveal the
nature of the phases.

The intermetallic compound V$_{3}$Si is a type-II superconductor that has
been known since the 1950s\cite{Hardy}. Stoichiometric and
near-stoichiometric samples form in the A15 crystal structure. The crystals
used in the present work were prepared using an electron-beam floating zone
technique \cite{Callaghan}. They were characterized by metallographic and
lattice parameter measurements. Also their resistivity ratios,
superconducting transition temperatures, and structural transition
temperatures were cataloged. The samples were flat wafers cut from the V$
_{3} $Si rods produced in the floating zone process, and polished using
standard abrasives. Superconducting transition temperatures ranged from
16.52 K to 16.97 K, and resistivity ratios from 2.6, for the lowest
transition temperature sample, up to 30 to 40 for samples with transition
temperatures the order of 16.95 K. Lattice parameters ranged from 4.7300
\AA\ for disordered samples down to 4.7242 \AA\ for the most ordered ones.
Although some wafers were single crystals, most of them were polycrystalline
with large crystallites.

Measurements of magnetization were made using a Quantum Design Magnetic
Properties Measurement System (MPMS). Samples, with their flat sections
parallel to the magnetic field direction, were attached to a stepper-motor
controlled platform and driven through a superconducting detection coil,
which is a second-order gradiometer connected to a superconducting quantum
interference device (SQUID). Standard software was used to determine the
magnetic moment from the SQUID output signal. Although this is a
conventional system for the determination of the properties of
superconducting and magnetic materials, some detail is relevant, as the
on-axis field homogeneity of the particular system (serial \#1 of the MPMS
instruments) is only 4 parts in 10,000 over a distance of 3 cm. This means
that during a 3 cm scan at a field of 15 kOe, the sample in effect will be
subject to a variation of magnetic field of order 6 Oe. Much has been
written about the effect on the reversibility of measurements of field
inhomogeneity in this instrument \cite{Suenaga,Ravikumar}. We do not believe
that much of this is relevant for this particular set of studies with these
samples as the m vs. H curves that were obtained were reversible under most
conditions. It is likely that the variation of the field during a scan is
too small a fraction of the total field to introduce any significant
additional irreversibility.

\begin{figure}[tbp]
{\epsfig{figure=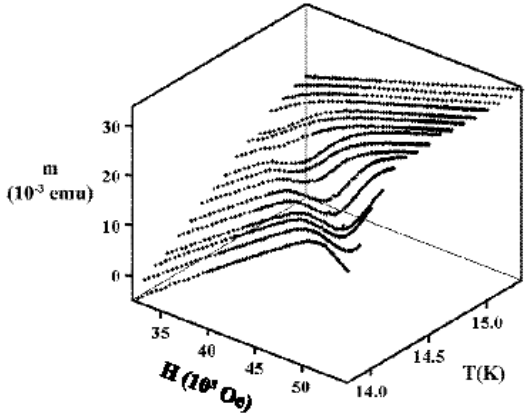,width=.45\textwidth}}
\caption{Three dimensional
plot showing the magnetic moment measured at fixed temperature, as a function of
applied magnetic field and temperature. The temperature ranges from 13.8 K to
15.2 K, in 0.1 K increments. }
\label{fig1}
\end{figure}

Standard plots of the magnetic moment vs. magnetic field, $m(H)$, for these V%
$_{3}$Si wafers were consistent with the usual results for type II
superconductors, with $H_{c1}$ measured to be 700 Oe. Because the upper
limit on the magnetic field in this apparatus was 55 kOe, and the upper
critical field of ${\rm V_{3}Si}$ at low temperatures is 200 kOe, there was a
relatively high lower bound on the temperature at which a complete set of
data could be obtained. In Fig.~\ref{fig1} we show a three-dimensional plot of raw
data of $m(H)$ at different fixed temperatures, for a temperature regime
relatively close to the zero-field critical temperature. This particular
sample was a wafer that was 0.76 cm x 0.48 cm in area and 0.053 cm thick.
Below a characteristic temperature, flat sections appear in $m(H)$ isotherms,
and as temperature is decreased further these deform into structures which
appear to oscillate reversibly (without hysteresis). The flat and
oscillatory regimes separate regions which are distinguished by different
values of the susceptibility, $\left( \partial m/\partial H\right) _{T}$. In
Fig.~\ref{fig2} we show $m(T)$ at fixed $H$, for a range of magnetic fields spanning
the regime of $T$ and $H$ in which there are nontrivial features in the $m(H)
$ isotherms. For this mode of data acquisition, hysteretic behavior in
temperature is clearly observed. This occurs even though there is no
apparent hysteresis in the $m(H)$ isotherms over the same range of
temperatures.

The above-described behavior was observed in five out of seven V$_{3}$Si
samples. It was always seen if the samples are cooled in zero field to low
temperatures, and the field and $m(H)$ curves are obtained with increasing
temperature. On the other hand, reversibility was lost in most of the
samples if they were cooled in high fields before a measurement cycle was
initiated. Even in this circumstance curves of $m(H)$ in increasing field
were the same as those obtained on cooling in zero field.

\begin{figure}[tbp]
{\epsfig{figure=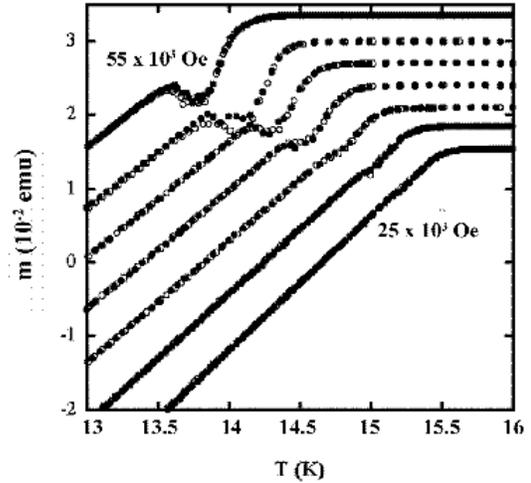,width=.45\textwidth}}
\caption{Magnetic moment as a function of temperature at different applied
fields: 25 (bottom), 30, 35, 40, 45, 50, and 55 x 10$^{3}$ Oe (top).
Closed symbols represent data taken as the temperature is
increased, while open symbols represent decreasing temperature data.
Noticeable hysteresis is observed as the field is increased beyond 35 x 10$
^{3}$ Oe. }
\label{fig2}
\end{figure}

It is apparent from the $m(H)$ isotherms shown in Fig.~\ref{fig1} that the
thermodynamics of the system may best be described by considering the
magnetic work $dW$ done by the system when $H$ is varied at constant $m$, $
dW=-mdH$ \cite{Kittel}. This also corresponds to the choice in which the
internal energy does not include the field energy. The magnetic moment, $m$,
in this instance is analogous to the pressure $P$ whereas the magnetic field, 
$H$, is analogous to the volume $V$. The variables $P$ and $V$ are used to
describe mechanical work in liquids. Thus $m(H)$ isotherms are analogs of $
P(V)$ isotherms. The data exhibit features that resemble those found in
liquid-gas transitions. There are isotherms in which there are regimes in
which $m$ is independent of $H$. As the temperature is lowered, the ``flat''
region is replaced by a regime in which $m$ rises and falls. This is
reminiscent of the behavior of the van der Waals equation for a fluid, which
can describe metastable persistence of the low and high density regimes,
into what is usually a two-phase region, when the pressure is constant over
a range of volumes. Only in this instance the magnetic moment appears to
move smoothly and reversibly between the phases, without abrupt nucleation
phenomena. Perhaps the appearance, with decreasing temperature, of a
continuous transformation between the two states, has to do with the growth
of barriers to changing the vortex configuration and moving vortices.

The existence of flat isotherms and the analogy with van der Waals isotherms
for the oscillatory isotherms suggests a two-phase regime with an endpoint 
\cite{Kierfield}. The analogy with the fluid system suggests that there is a
first order transition line terminating at a critical point. The two phases
associated with the first order phase transition would have to have the same
symmetry\cite{Lifshitz}. 
In Fig.~\ref{fig4} we show a phase diagram for the behavior of this sample.
The figure 
includes $H_{c2}(T)$, in addition to lines denoting the high and low-field
limits of the two-phase regime, delineated by the shaded region.
Because of the limitation to 55 kOe on the highest
field achievable with this apparatus, we were not able to explore this
behavior to lower temperatures than shown in Fig.~\ref{fig1}. As a consequence we do
not know whether the two-phase regime terminates at a second critical point
as the temperature is reduced, or persists to arbitrarily low temperatures.

\begin{figure}[tbp]
{\epsfig{figure=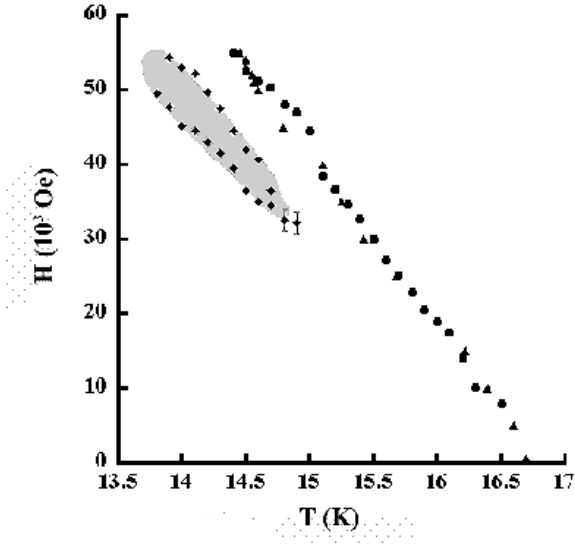,width=.45\textwidth}}
\caption[The phase diagram, magnetic field as a function of temperature.]{%
Phase diagram in $H-T$ space. The upper symbols denote $H_{c2}(T)$. The shaded
area 
denotes the putative two phase region
bounded by the coexistence curve (diamonds). It is not known whether this
region extends to low temperatures because of the limitation of magnetic
field in the measuring instrument. }
\label{fig4}
\end{figure}

The results discussed here are very different from those that have been
reported previously for V$_{3}$Si \cite{Isino,Chaudhary}. Their $m(H)$
isotherms were always hysteretic, exhibiting peaks and dips with decreasing
and increasing magnetic field, respectively. Similar behavior was found in
our most disordered samples, in our thick samples, and in our less
disordered samples when their flat parts were rotated such that they were at
an angle with respect to the field direction. The details of $m(H)$ curves
would thus appear to depend critically on the geometry of measurement, and
the degree of disorder, although a detailed mechanism is not known. (The
fact that the samples in which these effects were seen were very thin wafers
may account for our results as this might minimize demagnetizing effects as
well as the role of off-axis field inhomogeneities.) We must emphasize that
our samples also have dislocations and defects, and are usually not single
crystals, so our observations cannot be a consequence of having samples of
``superior'' quality. We believe that some combination of defects and
geometry allow a two-phase equilibrium state to be achieved for some, but
not for the full range of parameters.

There have been many discussions of nonlinear features of type-II
superconductors. In fact there has been somewhat of a debate as to whether
in paramagnetic materials such as V$_{3}$Si these features are a signature
of a generalized Fulde-Ferrell- Larkin-Ovchinnikov (GFFLO) phase \cite
{Tachiki} which can appear at high rather than at low temperatures as
originally predicted. In this phase, the superconductivity becomes spatially
inhomogeneous, with a wavelength that diverges at the transition to the
normal state. The transition to the GFFLO phase with increasing temperature
is predicted to be first order. There are no obvious features of the data
that would support this interpretation.

An alternative explanation is that our observations are associated with the
peak effect, a phenomenon known for many decades in conventional
superconductors \cite{Pippard}. In the peak effect, the critical current,
instead of decreasing as the superconductor-normal phase boundary is
approached, increases before falling to zero at the transition. The
reversibility of our $m(H)$ isotherms would seem to rule out our
observations as being attributable to dynamical effects associated with the
critical current. Our previous discussion points to two distinct {\it %
equilibrium} vortex phases, possibly glass phases, or disordered lattice
phases separated by a first order transition, which terminates at a critical
point. It is conceivable that in a dynamical regime these phases could have
different pinning currents as reported in Ref.~\cite{Marchevsky}.

An important issue is the question of what distinguishes the two phases
whose thermodynamic signature is inferred in our interpretation of $m(H)$
isotherms. The high- and low-field regimes on either side of the
``two-phase'' regime have different magnetic susceptibilities $\left(
\partial m/\partial H\right) _{T}$ and their vortex densities are different.
Thus the two coexisting phases would have different degrees of screening as $
m$ is fixed with $H$ changing, so that $B$ would have to change. In
equilibrium the free energies of the two phases at constant $T$ and $M$
would be the same. With our choice of dependent and independent variables,
this leads to a Clausius-Clapeyron equation \cite{Welp} of the form $%
dm/dT=\Delta S/\Delta H$.

A thermodynamic consistency check of a picture in which the two phases are
distinguished by their vortex densities is possible. One can determine $
\Delta S$ from measured quantities and compare this value with an estimate
obtained from the change in the number of states associated with the change
in the vortex density across the transition. To obtain $\Delta S$ from the
isotherms, we first estimate $dm/dT$ from the data of Fig.~\ref{fig1}. (For the
oscillatory isotherms we use the average moment.) We find $dm/dT$ to be
0.005 emu/K, and  relatively constant over the accessible range of
temperatures. The change in field across the ``flat'' region is about 3000
Oe. Using these values $\Delta S$ is found to be approximately $
10^{17}k_{B}. $

Then, the entropy per configuration in the vortex state, $S/k_{B}$, is given
by $\ln[N_{e}]$, where the number of configurations $N_{e}\approx N_{v}N_{c}$
. Here $N_{v}$ is the number of vortices threading the sample, and $N_{c}$
the number of configurations of a single vortex. The number of vortices $
N_{v}$ is given by the flux divided by the flux quantum. Taking the relevant
area of the crystal to be 0.05 x 0.5 cm$^{2}$, and the average magnetic
field to be 40,000 Oe, we find $N_{v}$ = 5 x 10$^{9}$ vortices. The number $
N_{c}$ can be estimated as the number of vibrational modes of a vortex line.
Thus $N_{c}$ is of the order of the number of unit cell layers, which with a
sample of length 0.7 cm, is 1.4 x 10$^{7}$ unit cell layers. Then the change
in entropy associated with the change in field between values 1 and 2 is
given by $S_{2}-S_{1}\approx k_{B}\left[ N_{v2}-N_{v1}\right] N_{c}\ln
\left( N_{av}N_{c}\right) .$ Here $N_{v2}$ and $N_{v1}$ are the numbers of
vortices associated with the two fields, and $N_{av}$ is the average number
of vortices in the two-phase regime. With an average field of 40,000 Oe, and
a field change of 3,000 Oe in the case of the data in question, $S_{2}-S_{1}$
is found to be $\approx $ 2.2 x 10$^{17}$k$_{B}$ in remarkable agreement
with the value estimated from the data. If the coherence length, $\xi \left(
0\right) ,$ estimated from $dH_{c2}/dT$ to be approximately 27\AA ,
determines the cutoff in the number of modes, this estimate would be a
factor of five lower. Estimates of the entropy change taking into account
only the entropy of the vortex cores yield a somewhat smaller estimate of
the entropy change.

In summary, magnetic moment -magnetic field isotherms of crystals of
superconducting V$_{3}$Si have been found to resemble the pressure-volume
isotherms of the liquid-gas transition. We have suggested that this is
evidence of a first-order phase transition and a two-phase regime, with a
critical point in the superconducting phase diagram. As mentioned, in this
instance, the two phases would have to possess the same symmetry. The
precise nature of vortex phases in the presence of weak disorder is quite
complicated \cite{Giamarchi}, so that further studies such as measurements
of specific heat, ac magnetic susceptibility, and small angle neutron
scattering would be needed to resolve this issue. This
work was supported in part by the
National Science Foundation under grant NSF/DMR- 0138209.

% now the references. delete or change fake bibitem. delete next three
%   lines and directly read in your .bbl file if you use bibtex.

% now the references. delete or change fake bibitem. delete next three
%   lines and directly read in your .bbl file if you use bibtex.

%figures follow here
%Fill in the caption in the braces of the \caption{} command. Put the label
%\begin{figure}
% \caption{}
% \label{}
% \end{figure}
%Here is an example of the general form of a table:
%that you will use with \ref{} command in the braces of the \label{} command.
%Insert the column specifiers (l, r, c, d, etc.) in the empty braces of the
%\begin{tabular}{} command.
% \begin{table}
% \caption{}
% \label{}
%\begin{tabular}{}
% \end{tabular}
% \end{table}
\end{twocolumn}

\begin{references}
\bibitem{Crabtree}  G. Crabtree and D. Nelson, Physics Today{\bf \ 50}, 38
(April 1997); {\it Defects and Geometry in Condensed Matter Physics}, by David
R. Nelson, Cambridge University Press, Cambridge, (2002).

\bibitem{Larkin}  For a review, see: G. Blatter, M. V. Feigel'man, V. B.
Geshkenbein, A. I. Larkin, and V. M. Vinokur, Rev. Mod. Phys. {\bf 66}, 1125
(1994).

\bibitem{Ling}  X. S. Ling, S. R. Park, B. A. McClain, S. M. Choi, D. C.
Denber, J. W. Lynn, Phys. Rev. Lett. {\bf 86}, 712 (2001).

\bibitem{Pippard}  A. B. Pippard, Phil. Mag. {\bf 19}, 217 (1969).

\bibitem{Marchevsky}  M. Marchevsky, M. J. Higgins, and S. Bhattacharya,
Nature {\bf 409}, 591 (2001).

\bibitem{Hardy}  G. F. Hardy and J. K. Hulm, Phys. Rev. {\bf 89,} 884
(1953); G. F. Hardy and J. K. Hulm, Phys. Rev. {\bf 93}, 1004 (1954).

\bibitem{Callaghan}  T. Callaghan, J. Schwanebeck, L. E. Toth, M. Dayan and
A. M. Goldman, J. Appl. Phys., {\bf 49}, 2523 (1978); Meichao Hu Chiang,
Masters Thesis, University of Minnesota, 1978, unpublished.

\bibitem{Suenaga}  M. Suenaga, D. O. Welch, and R. Budhani, Supercond. Sci.
Technol. {\bf 5}, S1 (1992).

\bibitem{Ravikumar}  G. Ravikumar, et al., Physica C {\bf 298}, 122 (1998)
and references cited therein.

\bibitem{Kittel}  For a physical discussion of the difference between $MdH$
and $HdM$ work terms, see: {\it Elementary Statistical Physics}, by C. Kittel,
John Wiley and Sons, Inc, New York (1958).

\bibitem{Kierfield}  Jan Kierfield and Valerii Vinokur, Phys. Rev. B {\bf 61}
, R14928 (2000).

\bibitem{Lifshitz}  See: {\it Statistical Physics, Part 1},  by E.
M. Lifshitz and L. P. Pitaevskii, Pergamon Press, Oxford (1980), p. 257.

\bibitem{Isino}  M. Isino, T. Kobayashi, N. Toyota, T. Fukase, and Y. Muto
Phys Rev. B{\bf 38}, 4457 (1988).

\bibitem{Chaudhary}  S. Chaudhary, A. K. Rajarajan, K. J. Singh, S. B. Roy,
and P. Chaddah, Physica C{\bf 353}, 29 (2001).

\bibitem{Tachiki}  M. Tachiki, {\it et al}., Z. Phys. B 100, 369-380 (1996);
P. Gegenwart, P. {\it et al}., Ann. Physik{\bf \ 5}, 307 (1996).

\bibitem{Welp}  U. Welp, J. Fendrich, W. K. Kwok, G. W. Crabtree, and B. W.
Veal, Phys. Rev. Lett. {\bf 76}, 4809 (1996).

\bibitem{Giamarchi}  T. Giamarchi and P. Le Doussal, Phys. Rev. B {\bf 52},
1242 (1995).


\end{references}
\end{document}